# Unraveling the Dynamic Importance of County-level Features in Trajectory of COVID-19

Qingchun Li[1*], Yang Yang[2], Wanqiu Wang[3], Sanghyeon Lee[4], Xin Xiao[5], Xinyu Gao[6], Bora Oztekin[7], Chao Fan[8], Ali Mostafavi[9]

[1] Ph.D. student, Zachry Department of Civil and Environmental Engineering, Texas A&M University, 199 Spence St., College Station, TX 77840; e-mail: qingchunlea@tamu.edu
[2] Master student, Department of Computer Science and Engineering, Texas A&M University, 199 Spence St., College Station, TX 77843-3112; e-mail: yangyangsandy@tamu.edu
[3] Master student, Department of Computer Science and Engineering, Texas A&M University, 199 Spence St., College Station, TX 77843-3112; e-mail: wanqiu.wang@tamu.edu
[4] Master student, Department of Computer Science and Engineering, Texas A&M University, 199 Spence St., College Station, TX 77843-3112; e-mail: mailto:sanghyeonlee@tamu.edu,
[5] Master student, Department of Computer Science and Engineering, Texas A&M University, 199 Spence St., College Station, TX 77843-3112; e-mail: xyx56@tamu.edu,
[6] Master student, Department of Computer Science and Engineering, Texas A&M University, 199 Spence St., College Station, TX 77843-3112; e-mail: xy.gao@tamu.edu,
[7] Master student, Department of Computer Science and Engineering, Texas A&M University, 199 Spence St., College Station, TX 77843-3112; e-mail: bora@tamu.edu
[8] Ph.D. Candidate, Department of Computer Science and Engineering, Texas A&M University, 199 Spence St., College Station, TX 77843-3112; e-mail: mailto:chfan@tamu.edu,
[9] Associate Professor, Zachry Department of Civil and Environmental Engineering, Texas A&M University, 199 Spence St., College Station, TX 77840; e-mail: amostafavi@civil.tamu.edu

**Abstract**

The objective of this study was to investigate the importance of multiple county-level features in the trajectory of COVID-19. We examined feature importance across 2,787 counties in the United States using data-driven machine learning models. Existing mathematical models of disease spread usually focus on case prediction with different infection rates without incorporating multiple heterogeneous features that could impact the spatial and temporal trajectory of COVID-19 in different stages. Recognizing this, we trained random forest models using 23 features representing six key influencing factors affecting pandemic spread: social demographics of counties, population activities, mobility within the counties, movement across counties, disease attributes, and social network structure. Also, we categorized counties into multiple groups according to their population densities, and we divided the trajectory of COVID-19 into three stages: the outbreak stage, the social distancing stage, and the reopening stage. The study aims to answer two research questions: (1) The extent to which the importance of heterogeneous features evolves in different stages; (2) The extent to which the importance of heterogeneous features varies across counties with different characteristics. We fitted a set of random forest models to determine weekly feature importance. The results showed that: (1) Social demographic features, such as gross domestic product, population density, and minority status maintained high-importance features throughout stages of COVID-19 across the 2787 studied counties; (2) Within-county mobility features had the highest importance in county clusters with higher population densities; (3) The feature reflecting the social network structure (Facebook, social connectedness index), had higher importance in the models for counties with higher population densities. The results show that the data-driven machine learning models could provide important insights to inform policymakers regarding feature importance for counties with various population densities and in different stages of a pandemic life cycle.

**Introduction**

COVID-19 has inflicted great loss in both economic and social dimensions. To inform pandemic situations, several studies developed mathematic models to predict the outbreak and trajectory of pandemics to provide important insights for use by policymakers for developing pandemic control measures. For example, Tizzoni et al. [1] predicted the epidemic spread among three European countries using a susceptible-infectious-recovered (SIR) model that accounted for human mobility. Balcan et al. [2] used the mathematical model to study global pandemic dynamics considering long-range and short-range mobility patterns. Ferguson et al. [3] developed the transmission model to test the effects of epidemic control measures (e.g., social distancing) with different virus reproduction rates. Wang et al. [4] developed a deep learning model to predict epidemic transmission using within-season and between-season observations as features. For the COVID-19 pandemic, many studies developed mathematical models to evaluate containment measures and to predict the potential outbreak and trajectory of COVID-19. Anastassopoulou et al. predicted the evolution of the spread of COVID-19 based on a Susceptible-Infectious-Recovered-Deceased (SIRD) model and the estimation of the reproduction number ($R_0$) [5]. Block et al. [6] simulated community spread scenarios of COVID-19 based on social network structures. Chang et al. [7] integrated the standard Susceptible-Exposed-Infectious-Recovered (SEIR) model with the origin-destination mobility networks to fit the trajectory of COVID-19 and predicted infections in the reopening stage. Gatto et al. [8] also used the standard SEIR model to evaluate the effectiveness of COVID-19 containment measures (such as mobility reduction and social distancing). Cintia et al. [9] used a regression model to investigate the relationship between human mobility and the transmission of COVID-19 in Italy. Perc et al. [10] used a simple iteration that relied only on confirmed cases to forecast the spread of COVID-19 in the United States, Slovenia, Iran, and Germany. Petropoulos and Makridakis [11] implemented a simple time-series forecasting to predict the spread of COVID-19 based on the data of confirmed cases, deaths and recoveries. Tomar and Gupta [12] used the LSTM model to predict the spread of COVID-19 in India and discussed the effectiveness of pandemic control measures, such as social distancing and lockdown. Chimmula and Zhang [13] also developed Long Short-term Memory (LSTM) deep learning model to predict the transmission of COVID-19 in Canada. The existing mathematical models seek to predict the trajectory of epidemics/pandemics based on a limited number of features, such as mobility patterns, reproduction rates of virus, observations within and between seasons, number of confirmed cases, deaths and recoveries. Most of the existing mathematical models, however, could account for only a limited number of features and could not simultaneously examine the importance of heterogeneous features, such as social demographics, population activities, mobility patterns, disease-related attributes, and social network structure-based various datasets.

Various studies related to COVID-19 have highlighted multiple influencing factors that would affect the pandemic spread. Dowd et al. [14] highlighted the importance of social and demographics attributes (mainly focusing on age structures of populations) affecting infection rates in populations. Nepomuceno et al. [15] found that other demographic factors could affect the spread of COVID-19. Multiple studies have reported the effects of population density [16,17], household size and composition, hygienic and sanitary conditions, access to healthcare services, case notification systems, and economic disparities [18] on the trajectory of the COVID-19 infections. Yancy [19], Dyer [20], Laurencin [21], and Millett et al. [22] pointed out the racial and ethnic disparities of COVID-19 that hit minorities harder. In addition to the social and demographic factors, additional studies have reported the role of population activities, such as visits to points of interests (e.g., hospitals, restaurants and recreation centers) [23–25] and staying at home [26–28] as they affect transmission risks of COVID-19. Kraemer et al. [29], Badr et al. [30], Jia et al. [31], Linka et al. [32] and Askitas et al. [33] investigated the extent to which mobility patterns would affect the spread of COVID-19. Liu et al. [34], Zhang et al. [35], You et al. [36], and Shim et al. [37] examined the effects of disease attributes, such as the reproduction number, $R_0$, on the trajectory of COVID-19. Furthermore, Bucur [38], Block et al. [6], and Kuchler et al. [39] discussed how the social network structures would affect the spread of COVID-19 in communities. A recent deep learning model proposed by Ramchandani et al. [40] accounted for several heterogeneous features (e.g., social demographics, population activities, and mobility pattern). While the existing studies inform about various heterogenous features affecting the trajectory of COVID-19 spread, limited knowledge exists about the importance of these features across different cities and communities and at different stages of the pandemic spread. For example, mobility restriction orders may have greater effect on counties with higher population densities [41], and long-distance mobility restriction orders are more effective in the outbreak stage of the epidemic, while local control measures, such as shelter-in-place and social distancing orders, would be more effective after the outbreak stage [29]. Unraveling the importance of various features in the trajectory of pandemics is a critical element for predictive surveillance and data-driven policy formulation. Hence, to address this important knowledge gap, we aim to answer two research questions in this paper: (1) The extent to which the importance of heterogeneous features evolved in different stages (e.g., outbreak stage, social distancing stage, and reopening stage); (2) The extent to which the importance of heterogeneous features varied across counties with different characteristics.

To answer these two research questions, we examined 23 features related to social demographics, population activities, mobility, social network structure, and disease-related attributes. We collected features of 2787 counties from March 24 to June 23, 2020, during the outbreak stage, the social distancing stage, and the reopening stage of COVID-19 pandemic. Table 1 provides a summary of the examined features and their underlying data. (Refer to the supplemental materials for the elaboration of each feature and of data sources.) Based on these features, we built a set of data-driven random forest models to study the dynamics of importance of each feature during different stages of COVID-19, and we evaluated the importance across counties with different population densities. We would like to note that the purpose of this paper is not to build a state-of-the-art prediction model of COVID-19, although the data-driven model could be a complementary tool to prediction models.

**Table 1. Collected Features for the Data-driven Model**

| **Social demographic features** | Population density | Constant feature |
|---|---|---|
| | Gross domestic product (GDP) | Constant feature |
| | Socioeconomic status | Constant feature |
| | Household composition and disability | Constant feature |
| | Minority status and language | Constant feature |
| | Housing type and transportation | Constant feature |
| | Epidemiologic factors | Constant feature |
| | Healthcare system factors | Constant feature |
| | Overall COVID-19 community Vulnerability Index (CCVI) | Constant feature |
| **Population activity features** | Point-of-interest visits | Time-dependent feature |
| | Social distancing index (SDI) | Time-dependent feature |
| | Urban activity index (social) | Time-dependent feature |

|  | Urban activity index (work) | Time-dependent feature |
|  | Urban activity index (traffic) | Time-dependent feature |
|  | Urban activity index (home) | Time-dependent feature |
|  | Venables Distance | Time-dependent feature |
| **Within-county mobility features** | Cuebiq county mobility index (CMI) | Time-dependent feature |
|  | Cuebiq shelter-in-place index (SIP) | Time-dependent feature |
| **Across-county mobility features** | County in-degree centrality | Time-dependent feature |
|  | County out-degree centrality | Time-dependent feature |
|  | Colocation degree centrality | Time-dependent feature |
| **Disease attribute feature** | Reproduction number ($R_0$) | Time-dependent feature |
| **Social network structure feature** | Social connectedness index | Constant feature |

## Methodology

For each week from March 24, 2020, to June 23, 2020, we created five random forest classifier models with ten-fold cross validation using 23 county-level features. Each model was trained and tested for each week in the studied period to investigate the evolution of feature importance during the COVID-19 pandemic. Different random forest models tend to divide counties into clusters based on population densities. Population density is shown to be a dominant feature affecting the number of infected cases. To examine the effects of features other than population density, we created models for clusters of counties with varying population densities. Fig. 1 illustrates included features and counties in five sets of random forest models. Fig. 2 illustrates clusters of counties with varying population densities in models. We explain each random forest model in the following sub-sections.

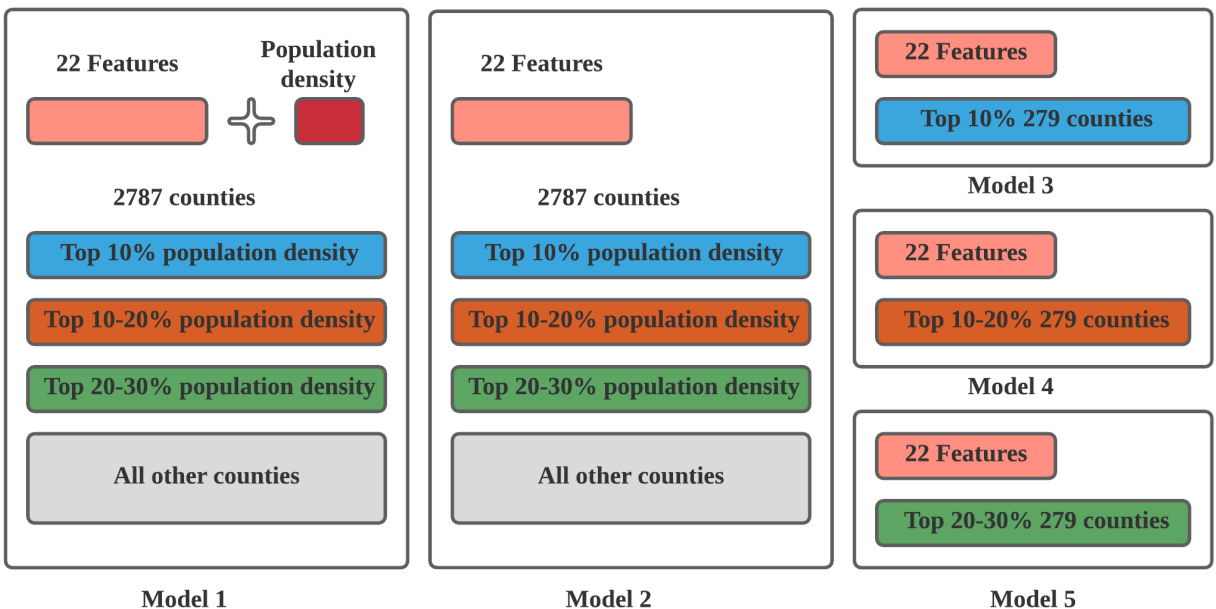

**Fig. 1** Features and counties included in five random forest models

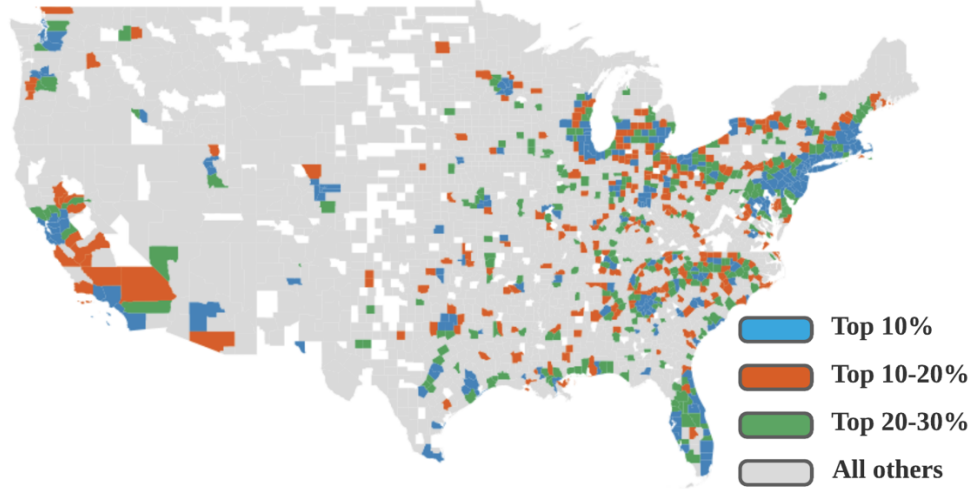

**Fig. 2** County population densities

*Two base models*

    We built two base random forest models. Model 1 includes all 23 features as independent variables; model 2 excludes population density, to enable assessment of the effect of other features in the absence of population density. The base models include 2787 counties in the United States. The dependent variable in each base model is composed of five classifications of the weekly new confirmed cases per 100,000 population (CPP) for each county provided by Centers for Disease Control and Prevention and Johns Hopkins University. For each week, we put counties with zero CPP in one classification. Then, we evenly divided the rest of counties into four classifications based on quantiles of CPP. Figure S1 in the supplemental materials illustrates histograms of the five classifications in each week. We used these two models as the baseline to compare feature importance in other models.

*Models of counties with different population densities*

    This set of random forest models include models 3, 4 and 5. Each model comprises 279 counties with top 10 (model 3), top 10 to top 20 (model 4), and top 20 to top 30 (model 5) percent of population densities among the 2787 counties. The dependent variable in these models is the same as that in the base models (i.e., five classifications based on the CPP in each week). The independent variables in this set of models include 22 features, excluding population density of each county. We used this set of models to examine feature importance of counties across different population density clusters.

    We examined the results of feature importance of each weekly model. Random forest modeling uses aggregated decreases in Gini importance of features to determine feature importance based on Equations 1 through 5 [42,43]. A greater aggregated decrease in a feature implies that the feature is more important.

$$\Delta G(j) = G(j) - \frac{|L_j|}{|j|}G(L_j) - \frac{|R_j|}{|j|}G(R_j) \qquad (1)$$

where, $G(j)$ is the Gini importance calculated accord to Equation 2, and $j$ is the partition at node $j$, while $L_j$ and $R_j$ are the left and right child nodes of partition $j$, respectively.

$$G(j) = \sum_{i=1}^{C} p_i(1 - p_i) \qquad (2)$$

where, $C$ is the total number of classes while $p_i$ is the probability of a datapoint from $j$ in class $i$.

$$F_i = \sum_{j:\text{node } j \text{ splits on feature } i} \Delta G(j) \qquad (3)$$

$$\text{Norm } F_i = \frac{F_i}{\sum_{j \in \text{all features in one tree}} F_j} \qquad (4)$$

where, $F_i$ is the importance of feature $i$ in one decision tree and is normalized between 0 to 1 according to Equation 4. Then final feature importance in the forest is determined by averaging normalized feature importance of all the trees (Equation 5).

$$RF\ F_i = \frac{\sum_{j:\text{all trees}} \text{Norm } F_{ij}}{T} \qquad (5)$$

where, $T$ is the total number of trees in the forest and $\text{Norm } F_{ij}$ is the normalized importance of feature $i$ in tree $j$.

Note that previous studies showed that feature importance based on the reduction of Gini importance would perform worse for categorical features, and some studies proposed new algorithms to correct the bias [44–46]. In this paper, we still used the reduction of Gini importance due to the features in our models are all numerical features.

**Results**
We found disparate patterns of feature importance at different stages of the COVID-19 pandemic and across counties with different population densities. Fig. 3 and Fig. 4 illustrate the ranks of feature importance in model 1 and model 2, which include all 2,787 counties (Refer to the supplemental materials for information about the accuracy of the random forest models.) Note that some models have relatively low accuracy on the test dataset, especially models with 279 counties (models 3, 4, and 5) with different population densities. The selected Gini feature importance, however, is calculated based on the training dataset, which had high accuracies [42]. Therefore, the examined feature importance in this study could still provide insights regarding to what extent the features are important to classify and separate the counties based on CPP.

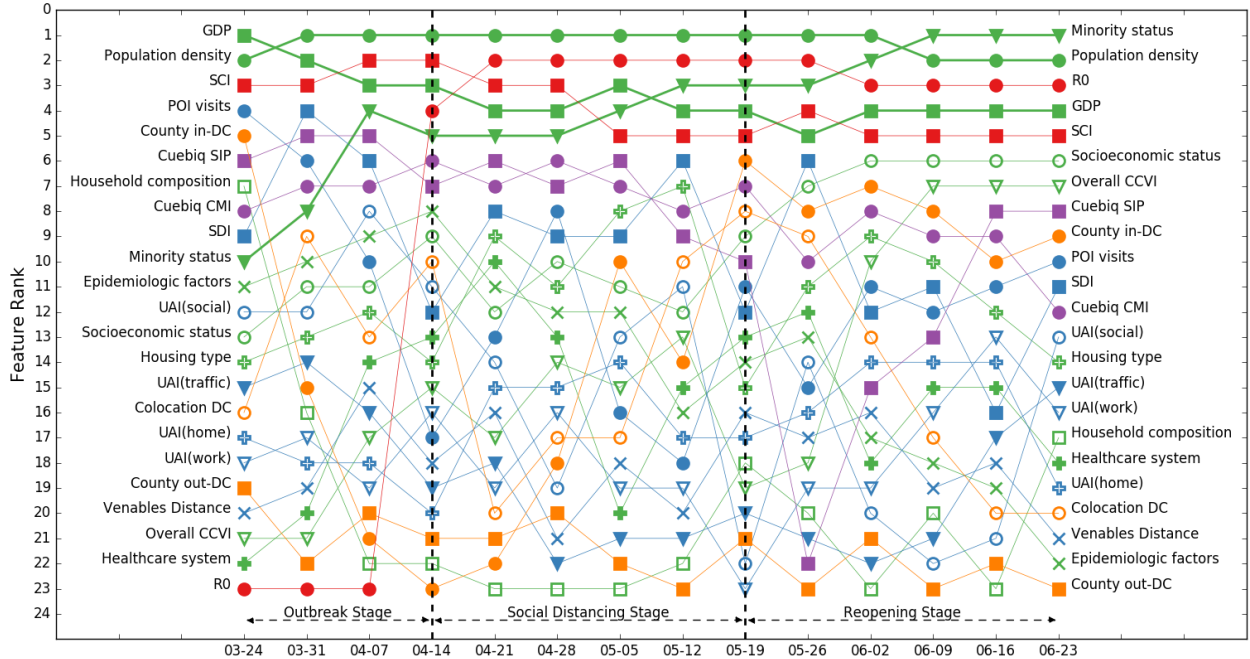

**Fig. 3** Rank of feature importance of model 1. Color coding of 23 features: green: social demographic features, blue: population activity features, purple: within-county mobility features, orange: across-mobility features, and red: disease attribute feature and social network structure feature. The legend also distinguishes each feature. Features with greater importance are highlighted with thicker lines.

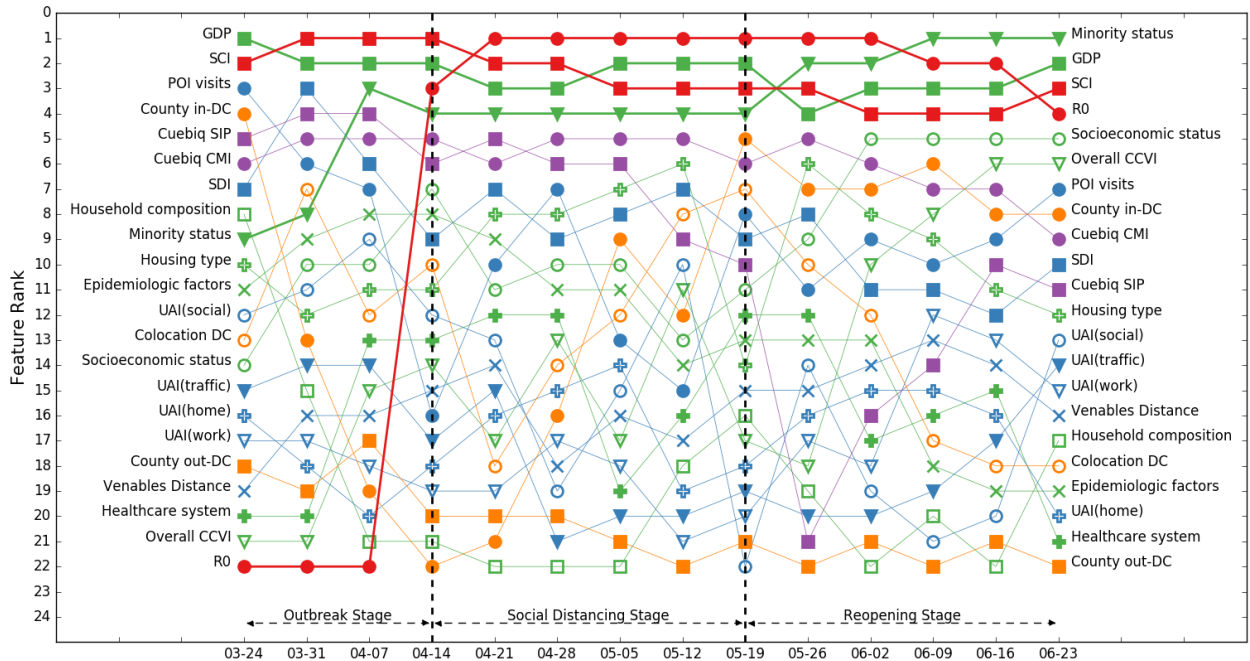

**Fig. 4** Rank of feature importance of model 2.

*Feature importance in models 3 and 4:*

We can observe from Fig. 3 and Fig. 4 that ranks of feature importance of models 1 and 2 are quite close. Some features retained high importance while some features showed varied importance in different stages of the COVID-19 pandemic.

A. Features keeping high importance across stages:

We can find that gross domestic product (GDP), population density, and social connectedness index (SCI) maintained high feature importance across all stages of COVID-19 in models 1 and 2, both of which include all 2787 counties. Population density stayed within the top two most important features, while GDP and SCI stayed in the top 5 most important features. Although we removed the influence of population size in the weekly confirmed cases—we considered confirmed cases per 100,000 population—the results of feature importance still indicate that population density has a significant impact on CPP. This is consistent with the results of existing studies that examined the effects of population density on the spread of COVID-19 [16,17]. GDP is another important social demographic variable. GDP would highly affect other social demographic variables, such as household size and composition, hygienic and sanitary conditions, access to healthcare services, case notification systems, and economic disparities. Existing research has highlighted the importance of these social demographic variables in the spread of the COVID-19 pandemic [15]. Furthermore, SCI captures the effects of social networks; the result indicates that social networks greatly affect the risks of virus spread during all stages of the pandemic. This result is consistent with existing studies that more connected social network structures are at a greater risk of virus spread during the pandemic [6,38,39].

B. Features showing increasing importance across stages:

We can observe that the importance of reproduction number ($R_0$), minority status, socioeconomic status, and COVID-19 community vulnerability index (CCVI), all of which showed increasing importance across stages.

$R_0$ was of low importance in the outbreak stage and spiked to the highest importance during the social distancing stage and remained in the top four most important features in the reopening stage. This result implies that $R_0$ is an important feature in determining the extent of disease spread when community spread begins. The importance of minority status showed a similar pattern: it was of relatively low importance in the outbreak stage, then rose to the top four in the social distancing stage and became the highest-importance feature in the reopening stage. This result supports the findings of other studies by Yancy [19], Dyer [20], Laurencin [21], and Millett et al. [22] that reported a greater exposure to the virus in racial minority populations. Another two social demographic features, socioeconomic status and CCVI, also showed relatively low importance in the outbreak stage, increasing importance in the social distancing stage, and relatively high importance in the reopening stage. The results not only highlight the importance of social demographic features in the spread of virus, but also shed light on the criticality of incorporating a functional timeline that takes into account relevant features when developing pandemic control policies. For example, reopening planning should account for populations with different socioeconomic status and CCVI because these two features demonstrate high importance in the reopening stage.

C. Features showing decreasing importance across stages:

We found that the importance of two within-county mobility features, Cuebiq county mobility index (CMI) and Cuebiq shelter-in-place index (SIP) showed an overall decreasing trend across stages in two base models. The importance of both was high in the outbreak stage. This result implies the importance of mobility reduction measures in the initial outbreak to slow down community spread [29]. The importance of CMI was also high in the social distancing stage and showed a decreasing trend in the reopening stage, while the importance of SIP showed a decreasing trend in the social distancing stage and became increasingly high in the reopening stage. These results could imply that counties which maintained their social distancing practices even after reopening were more likely to maintain or to decrease their number of infection cases.

D. Other highlighted features:

For features related to population activities, point-of-interest (POI) visits had high feature importance at the beginning of the outbreak stage, but its importance started waned in the following weeks, reaching lowest importance at the end of the outbreak stage. In the social distancing stage, the importance of POI visits fluctuated: it increased during the first two weeks then decreased in the following two weeks. In the reopening stage, the importance of POI visits showed an increasing trend. The result could imply the importance of maintaining the reduction in POIs visits in the social distancing and reopening stage for the containment of the virus spread. The social distancing index (SDI), although showing a fluctuating pattern, had a relatively high feature importance across all three stages. Similarly, this result supports the importance of voluntary and mandatory social distancing measures for virus spread reduction across all stages.

The importance of cross-county mobility features and county out-degree and in-degree centrality revealed important insights regarding travel reduction. County in-degree centrality, like POI visits, had relatively high importance at the beginning of the outbreak stage but showed a decreasing trend until the end of outbreak stage. In the social distancing stage, the importance of county in-degree centrality showed an increasing trend, reached its highest importance rank at the end of social distancing stage and kept relatively high feature importance in the reopening stage. The importance of county out-degree centrality, on the other hand, kept low across three stages of COVID-19. The results highlight the importance of travel reduction and limited cross-county movements in all stages of the pandemic, especially for the movements into counties. Counties should limit and monitor the number of travelers from other counties to effectively contain the spread of the virus from their counties.

*Feature Importance across Population Density Clusters*

In the next step, we investigated feature importance for counties with different population densities. Figures 5 through 7 illustrate models 3, 4 and 5, which encompass data from 279 counties with the top 10%, the top 10%–20%, and top 20%–30% densities among the 2787 counties.

**Fig. 5** Rank of feature importance of model 3

**Fig. 6** Rank of feature importance of model 4

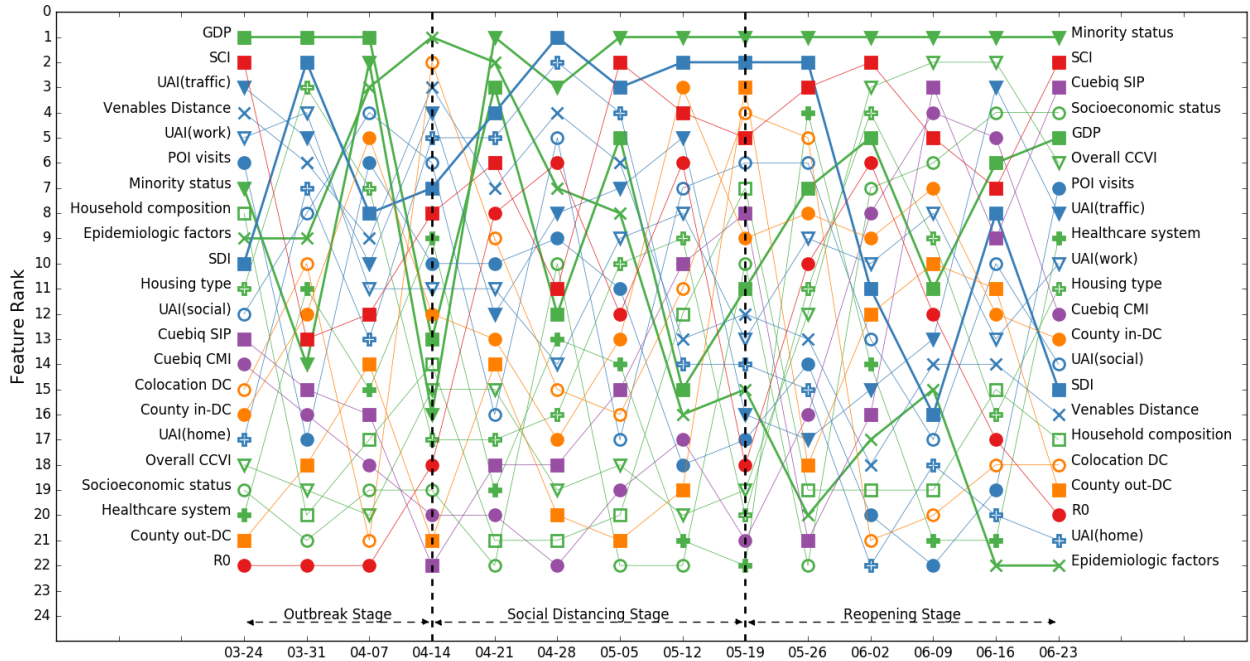

**Fig. 7** Rank of feature importance of model 5

We can observe from Figures 5 through 7 that feature importance shows different patterns across population density clusters. Some features showed high importance only in the clusters with high population density, and some features showed lower importance in population density clusters compared with importance of the two base models.

A. Features having high importance in the high-population density cluster:

We can observe that two within-county mobility features, Cuebiq CMI and SIP, showed high importance only in the top 10% population density cluster (model 3). For counties with lower population densities in models 4 and 5, the importance of these two within-county mobility features decreased, with the lowest importance in model 5. In model 3, these two features showed high importance in the outbreak and social distancing stages. In the reopening stage, the importance of SIP showed a decreasing trend, while CMI decreased in the first three weeks then showed an increasing trend again. The results are consistent with existing studies that indicate that mobility restriction orders could be more effective in counties with higher population densities [41]. Also, mobility restriction orders are more effective in early stage (i.e.. outbreak stage) of the epidemic, and local control measures, such as shelter-in-place and social distancing orders, would be more effective after the outbreak stage [29].

Also, another dominant feature in the base models, social connectedness index (SCI), had a pattern similar to that of Cuebiq CMI. In model 3, SCI showed high importance in the outbreak stage, and the importance started to decrease in the social distancing stage. In the reopening stage, the importance of SCI showed an increasing trend. The results imply that SCI is more important in the county clusters with higher population densities, and SCI had higher and increased importance when there were more and increased human interactions. Hence, policy makers should account for social connectedness in the area with higher population densities.

B. Features showing high importance in low population density clusters:

We found that social distance index had higher importance in low population density clusters. SDI represents the portion of the number of digital devices at home divided by the total number of digital devices in the area. In models 4 and 5 SDI, showed higher importance in the social distancing stage. In model 3, although SDI had lower importance compared with models 4 and 5, it still showed an increasing importance in the social distance stage. The results could imply that SDI is an important feature for effective pandemic control in the social distancing stage. Also, social distancing may more effectively help pandemic control in counties with low population densities in the social distancing stage. The results show that it is important to develop and monitor social distancing measures, especially for the counties with low population densities.

C. Features shoingd lower importance and features keeping high importance in population density clusters:

We found that in models 3 through5, GDP and $R_0$ did not show the same importance compared with the base models inclusive of all counties. This result may imply the correlation between GDP, $R_0$, and population density. The minority status feature, however, still showed high importance in models of population density clusters, especially in the social distancing and reopening stages. The results indicate that for the county clusters with close population densities, GDP and $R_0$ are not important features in examining the spread of virus. Policy makers should to racial minority groups, especially in the social distancing and reopening stage. Formulating policies that could help racial minority groups may effectively help with the overall pandemic controls.

**Discussion**

The majority of the existing literature on epidemic spread modeling and COVID-19 pandemic primarily focuses on standard epidemiological models for examining the effects of population features on disease spread. However, the ability of these models to examine the relative importance of various features across different stages of the disease spread based on various datasets is rather limited. To address this gap, in this paper, we investigated the importance of collected 23 heterogeneous features in the trajectory of COVID-19 using a data-driven machine-learning model comprising 2,787 counties in the United States. The results demonstrate the dynamics of feature importance among counties in United States and across three stages of the COVID-19 pandemic. In the models including all 2,787 counties, (1) social demographic features, such as GDP and population density, and the feature reflecting social interaction strength, social connectedness index, kept high importance through stages of the COVID-19 pandemics; (2) a virus attribute feature, reproduction number ($R_0$), and some social demographic features, including minority status, socioeconomic status, and COVID-19 community vulnerability index (CCVI) showed increased importance in the trajectory of the COVID-19 pandemic; (3) within-county mobility features, Cuebiq county mobility index (CMI) and shelter-in-place (SIP), showed decreased importance across different stages; while in the models with different population densities, the level of importance varied; (4) within-county mobility features showed higher importance in county clusters with higher population densities; (5) GDP and $R_0$ did not show the same importance within the models encompassing 2,787 counties, while the minority status feature still showed an initial low level and increasing importance across stages; and (6) social distance index (SDI) showed higher importance in county clusters with lower population densities and higher importance in the social distancing and reopening stages.

The results showed consistency with findings of other studies. For example, Nepomuceno et al. [15] highlighted that social demographic factors would greatly affect the spread of COVID-19. Also, Yancy [19], Dyer [20], Laurencin [21], and Millett et al. [22] argued that minority groups are disproportionately affected by COVID-19. Kuchler et al. [39] showed social network structure would affect the spread of COVID-19 among counties. Furthermore, the results showed that within-county mobility features had higher importance in the model of counties with higher population densities, which is consistent with the argument that mobility restriction orders may have greater effect on counties with higher population densities [41]. Our study, however, examined the importance of multiple heterogeneous features related to population activities, sociodemographic attributes, virus features, and mobility simultaneously. For example, our study revealed that the importance of social network structure decreased in the model of counties with lower population densities. Social distancing index had higher importance in the model of counties with lower population densities. Minority status kept high importance in different population density clusters.

The results could help policymakers develop pandemic control measures and strategies at different levels and at different timepoints. For example, when policymakers develop pandemic control measures at the country level, GDP, population densities, and SCI could be effective indictors to separate counties for customized pandemic control measures. When considering a specific county, such as New York County (also known as Manhattan) that has the highest population density, pandemic control measures could focus on within-county mobility and social network structure in the outbreak and social distancing stages, while minority status, socioeconomic status, and overall COVID-19 community vulnerability index could be employed in the social distancing and reopening stages. In smaller counties (such as Brazos County, Texas), pandemic control measures may need to focus on social distancing in the outbreak and social distancing stages, while in the reopening stage, social demographic factors, such as minority status and CCVI should be accounted for. The feature analysis importance could also suggest what feature indicators should be monitored by public officials in different counties and across different stages of a pandemic.

We would like to note some limitations in this study. We considered only first-order feature importance in this paper. Features may have interactions that deviate from an additive linear explanation at different timepoints and for models of counties with different characteristics. The dynamics of feature interaction in the trajectory of pandemics could be explored in future studies. The results of feature interaction could inform which feature will strengthen or weaken another feature. In this paper, we collected only 23 county level features. Future studies could explore and use more county-level features to complement these results.

**References**

1. Tizzoni, M. *et al.* On the Use of Human Mobility Proxies for Modeling Epidemics. *PLoS Comput. Biol.* **10**, (2014).
2. Balcan, D. *et al.* Multiscale mobility networks and the spatial spreading of infectious diseases. *Proc. Natl. Acad. Sci. U. S. A.* **106**, 21484–21489 (2009).
3. Ferguson, N. M. *et al.* Strategies for containing an emerging influenza pandemic in Southeast Asia. *Nature* **437**, 209–214 (2005).
4. Wang, L., Chen, J. & Marathe, M. DEFSI: Deep Learning Based Epidemic Forecasting with Synthetic Information. *Proc. AAAI Conf. Artif. Intell.* **33**, 9607–9612 (2019).
5. Anastassopoulou, C., Russo, L., Tsakris, A. & Siettos, C. Data-based analysis, modelling

and forecasting of the COVID-19 outbreak. *PLoS One* **15**, (2020).
6. Block, P. *et al.* Social network-based distancing strategies to flatten the COVID-19 curve in a post-lockdown world. *Nat. Hum. Behav.* **4**, 588–596 (2020).
7. Chang, S. *et al.* Mobility network models of COVID-19 explain inequities and inform reopening. *Nature* (2020). doi:10.1038/s41586-020-2923-3
8. Gatto, M. *et al.* Spread and dynamics of the COVID-19 epidemic in Italy: Effects of emergency containment measures. *Proc. Natl. Acad. Sci. U. S. A.* **117**, 10484–10491 (2020).
9. Cintia, P. THE RELATIONSHIP BETWEEN HUMAN MOBILITY AND VIRAL TRANSMISSIBILITY DURING THE COVID-19 EPIDEMICS IN ITALY. *arXiv e-prints* arXiv:2006.03141 (2020).
10. Perc, M., Gorišek Miksić, N., Slavinec, M. & Stožer, A. Forecasting COVID-19. *Front. Phys.* **8**, (2020).
11. Petropoulos, F. & Makridakis, S. Forecasting the novel coronavirus COVID-19. *PLoS One* **15**, (2020).
12. Tomar, A. & Gupta, N. Prediction for the spread of COVID-19 in India and effectiveness of preventive measures. *Sci. Total Environ.* **728**, (2020).
13. Chimmula, V. K. R. & Zhang, L. Time series forecasting of COVID-19 transmission in Canada using LSTM networks. *Chaos, Solitons and Fractals* **135**, (2020).
14. Dowd, J. B. *et al.* Demographic science aids in understanding the spread and fatality rates of COVID-19. *Proc. Natl. Acad. Sci. U. S. A.* **117**, 9696–9698 (2020).
15. Nepomuceno, M. R. *et al.* Besides population age structure, health and other demographic factors can contribute to understanding the COVID-19 burden. *Proceedings of the National Academy of Sciences of the United States of America* **117**, 13881–13883 (2020).
16. Rocklöv, J. & Sjödin, H. High population densities catalyse the spread of COVID-19. *Journal of travel medicine* **27**, (2020).
17. Ahmadi, M., Sharifi, A., Dorosti, S., Jafarzadeh Ghoushchi, S. & Ghanbari, N. Investigation of effective climatology parameters on COVID-19 outbreak in Iran. *Sci. Total Environ.* **729**, (2020).
18. Wright, A. L., Sonin, K., Driscoll, J. & Wilson, J. Poverty and Economic Dislocation Reduce Compliance with COVID-19 Shelter-in-Place Protocols. *SSRN Electron. J.* (2020). doi:10.2139/ssrn.3573637
19. Yancy, C. W. COVID-19 and African Americans. *JAMA - Journal of the American Medical Association* **323**, 1891–1892 (2020).
20. Dyer, O. Covid-19: Black people and other minorities are hardest hit in US. *BMJ* **369**, m1483 (2020).
21. Laurencin, C. T. & McClinton, A. The COVID-19 Pandemic: a Call to Action to Identify and Address Racial and Ethnic Disparities. *J. Racial Ethn. Heal. Disparities* **7**, 398–402 (2020).
22. Millett, G. A. *et al.* Assessing differential impacts of COVID-19 on black communities. *Ann. Epidemiol.* **47**, 37–44 (2020).
23. Benzell, S. G., Collis, A. & Nicolaides, C. Rationing social contact during the COVID-19 pandemic: Transmission risk and social benefits of US locations. *Proc. Natl. Acad. Sci.* 202008025 (2020). doi:10.1073/pnas.2008025117
24. Chang, S. Y. *et al.* Mobility network modeling explains higher SARS-CoV-2 infection rates among disadvantaged groups and informs reopening strategies. *medRxiv*

2020.06.15.20131979 (2020). doi:10.1101/2020.06.15.20131979
25. Bahl, P. *et al.* Airborne or Droplet Precautions for Health Workers Treating Coronavirus Disease 2019? *J. Infect. Dis.* (2020). doi:10.1093/infdis/jiaa189
26. Friedson, A., McNichols, D., Sabia, J. & Dave, D. Did California's Shelter-In-Place Order Work? Early Coronavirus-Related Public Health Effects. *Natl. Bur. Econ. Res.* (2020). doi:10.3386/w26992
27. Lyu, W. & Wehby, G. L. Shelter-In-Place Orders Reduced COVID-19 Mortality And Reduced The Rate Of Growth In Hospitalizations. *Health Aff. (Millwood).* (2020). doi:10.1377/hlthaff.2020.00719
28. Sen-Crowe, B., McKenney, M. & Elkbuli, A. Social distancing during the COVID-19 pandemic: Staying home save lives. *American Journal of Emergency Medicine* **38**, 1519–1520 (2020).
29. Kraemer, M. U. G. *et al.* The effect of human mobility and control measures on the COVID-19 epidemic in China. *Science (80-. ).* **368**, 493–497 (2020).
30. Badr, H. S. *et al.* Association between mobility patterns and COVID-19 transmission in the USA: a mathematical modelling study. *Lancet Infect. Dis.* (2020). doi:10.1016/S1473-3099(20)30553-3
31. Jia, J. S. *et al.* Population flow drives spatio-temporal distribution of COVID-19 in China. *Nature* **582**, 389–394 (2020).
32. Linka, K., Peirlinck, M., Sahli Costabal, F. & Kuhl, E. Outbreak dynamics of COVID-19 in Europe and the effect of travel restrictions. *Comput. Methods Biomech. Biomed. Engin.* (2020). doi:10.1080/10255842.2020.1759560
33. Askitas, N., Tatsiramos, K. & Verheyden, B. Lockdown Strategies, Mobility Patterns and COVID-19. *arXiv Prepr. arXiv2006.00531* (2020).
34. Liu, Y., Gayle, A. A., Wilder-Smith, A. & Rocklöv, J. The reproductive number of COVID-19 is higher compared to SARS coronavirus. *J. Travel Med.* **27**, (2020).
35. Zhang, S. *et al.* Estimation of the reproductive number of novel coronavirus (COVID-19) and the probable outbreak size on the Diamond Princess cruise ship: A data-driven analysis. *Int. J. Infect. Dis.* **93**, 201–204 (2020).
36. You, C. *et al.* Estimation of the time-varying reproduction number of COVID-19 outbreak in China. *Int. J. Hyg. Environ. Health* **228**, (2020).
37. Shim, E., Tariq, A., Choi, W., Lee, Y. & Chowell, G. Transmission potential and severity of COVID-19 in South Korea. *Int. J. Infect. Dis.* **93**, 339–344 (2020).
38. Bucur, D. & Holme, P. Beyond ranking nodes: Predicting epidemic outbreak sizes by network centralities. *PLoS Comput. Biol.* **16**, e1008052 (2020).
39. Kuchler, T., Russel, D. & Stroebel, J. The geographic spread of COVID-19 correlates with structure of social networks as measured by Facebook. *arXiv e-prints: 2004.03055* (2020).
40. Ramchandani, A., Fan, C. & Mostafavi, A. DeepCOVIDNet: An Interpretable Deep Learning Model for Predictive Surveillance of COVID-19 Using Heterogeneous Features and Their Interactions. *arXiv Prepr. arXiv2008.00115* (2020).
41. Engle, S., Stromme, J. & Zhou, A. Staying at Home: Mobility Effects of COVID-19. *SSRN Electron. J.* (2020). doi:10.2139/ssrn.3565703
42. Breiman, L. Random forests. *Mach. Learn.* **45**, 5–32 (2001).
43. Kelly, C. & Okada, K. Variable interaction measures with random forest classifiers. in *Proceedings - International Symposium on Biomedical Imaging* 154–157 (2012). doi:10.1109/ISBI.2012.6235507

44. Altmann, A., Toloşi, L., Sander, O. & Lengauer, T. Permutation importance: A corrected feature importance measure. *Bioinformatics* **26**, 1340–1347 (2010).
45. Wright, M. N., Dankowski, T. & Ziegler, A. Unbiased split variable selection for random survival forests using maximally selected rank statistics. *Stat. Med.* **36**, 1272–1284 (2017).
46. Nembrini, S., König, I. R. & Wright, M. N. The revival of the Gini importance? *Bioinformatics* **34**, 3711–3718 (2018).

**Acknowledgements**
The authors would like to acknowledge funding support from the National Science Foundation RAPID project #2026814: Urban Resilience to Health Emergencies: Revealing Latent Epidemic Spread Risks from Population Activity Fluctuations and Collective Sense-making. The authors would also like to acknowledge that SafeGraph provided points-of-interest data. Any opinions, findings, conclusions, or recommendations expressed in this research are those of the authors and do not necessarily reflect the view of the funding agency. The authors would like to thank Jan Gerston for her copy editing service.

**Author contributions**
Q.L. and A.M. designed the study. Q.L., Y.Y., W.W., S.L., X.X., X.G., B.O., C.F. collected data, performed analysis and visualized the results. Q.L. wrote the paper and all the authors reviewed the paper.

**Competing interests**
The authors declare no competing interests.

# SUPPLEMENTAL MATERIALS

## 1. Collected features for the data-driven random forest model

*Features related to social demographics:*

**Population density**
We calculated population density of each county (population/square miles), as previous works showed that population density is an important factor influencing the spread of an epidemic (Ahmadi et al. 2020; Rocklöv and Sjödin 2020). Population density data was calculated based on the county-level Social Vulnerability Index 2018 of United States published by Centers for Disease Control and Prevention (CDC) (2020).

**Gross domestic product (GDP)**
We included GDP of each county in 2018 as a feature for the model, as previous works showed that GDP could be a vulnerability index for COVID-19. Counties with higher GDP usually have a more robust economy and better health systems compared with counties with lower GDP (Nepomuceno et al. 2020; Sarmadi et al. 2020). We used the 2018 county-level GDP published by United States Department of Commerce (2020).

**COVID-19 Community Vulnerability Index**
This study incorporates the county-level COVID-19 Community Vulnerability Index developed by Surgo Foundation based on the CDC data (Surgo Foundation 2020) which comprises seven social and demographic features determined by previous studies to affect the spread of COVID-19. (Nepomuceno et al. 2020; Wright et al. 2020).

1. Socioeconomic status: A measure accounting for population education, income, and occupation. Surgo Foundation developed this feature based on CDC's Social Vulnerability Index, which accounts for population below poverty, unemployed, and without a high school diploma.
2. Household composition and disability: Developed based on CDC's SVI, this feature accounts for populations aged 65 or older, populations aged 17 or younger, populations older than 5 years of age with a disability, and single-parent households.
3. Minority status and language: Also based on CDC's SVI, this feature accounts for minority and populations who speak English "less than well."
4. Housing type and transportation: Based on CDC's SVI, this feature accounts for the population's housing types, such as multi-unit structures, mobile homes, and crowded housing. It also accounts for populations without vehicles and those who live in group quarters.
5. Epidemiologic factors: Developed by Surgo Foundation in response to COVID-19, this feature accounts for populations with underlying conditions (e.g., cardiovascular, respiratory, immunocompromised, obesity, and diabetes) that are vulnerable to COVID-19.
6. Healthcare system factors: Developed by Surgo Foundation for COVID-19, this factor accounts for poor health system capacity, strength, and preparedness.
7. Overall COVID-19 community vulnerability index (CCVI): This feature combines the above six features with equal weights. CCVID is a composite score that reflects the extent of a county's vulnerability to COVID-19.

*Features related to population activities*

**Points-of-interest visits**
We used SafeGraph data, Weekly Pattern Version 2 (SafeGraph 2020a), to calculate the number of visits to points of interest (POIs) in each county, such as restaurants, museums, hospitals, and colleges. Furthermore, to remove the influence of disparate numbers of POIs in each county, we used the percentage change based on baseline POI visits of the first week, the week of March 3, 2020.

**Social distancing index**
We used Social Distancing Metrics developed by SafeGraph (SafeGraph 2020b), to calculate the social distancing index (SDI) of each county. The SDI was calculated dividing the number of cell phones at home by the total number of devices within a county. Also, we used the percentage change based on SDI of the first week to remove the potential influence of disparate amounts of devices in each county.

**Urban activity index**
We used Mapbox data to calculate the urban activity index. Mapbox data provides contact activity metrics in pre-defined tiles (measuring about 100 by 100 meters square) in 4-hour temporal resolution. We classified tiles into four categories and calculated an aggregated contact activity metric in classified tiles using Equation 1 to assess four kinds of urban activities on a larger scale. Therefore, urban activity index includes four sub-features: social activity index, traffic activity index, work activity index, and home activity index.
1. Social tiles: We classified tiles as social tiles in areas where at least one POI in SafeGraph is located.
2. Traffic tiles: Traffic tiles includes tiles incorporating roads.
3. Home tiles: Home tiles include tiles that cover residual buildings or have device information from 7 p.m. to 3 a.m.
4. Work tiles: all other tiles.

$$R_g = \sqrt{\frac{1}{n} \times (c_1^2 + c_1^2 + \cdots + c_n^2)} \qquad (1)$$

where $c_i$ represents contact activity metric in the tile of index $i$.

**Venables distance**
We also used Mapbox data to calculate the daily Venables distance of each county (Louail et al. 2015) according to Equation 2, reflecting the concentration of population activities.

$$D_V(t) = \frac{\sum_{i<j} s_i(t) s_j(t) d_{ij}}{\sum_{i<j} s_i(t) s_j(t)} \qquad (2)$$

where $s_i(t)$ and $s_j(t)$ are the daily average activity intensities in cells $i$ and $j$, respectively, and $d_{ij}$ is the distance between two cells. The resolution of cell is 4 square meters. We also calculated percentage change based on the values of the first week to remove the influence of disparate tiles and cells in each county.

*Features related to mobility within counties*

**Cuebiq county mobility index (CMI)**
We used a county mobility index provided by Cuebiq as a feature of population mobility within counties (Cuebiq 2020). The Cuebiq mobility index of each county is the median of aggregated movements of each user in a day in the county (Equations 3 and 4). For example, a CMI of 5 for a county represents that the median user in that county travels 500 meters.

$$Device\ Mobility\ (DM) = log_{10}(D + 1) \quad (3)$$
$$County\ Mobility\ Index\ (CMI) = Median(DM) \quad (4)$$

where $D$ is the aggregated distance of one user in a day in the county.

**Cuebiq shelter-in-place index (SIP)**
We used shelter-in-place index (SIP) provided by Cuebiq as another feature of population mobility within counties (Cuebiq 2020). Cuebiq calculated the percentage of users in each county who traveled less than 330 feet as the SIP.

*Features related to mobility across counties*

**County in-degree and out-degree centrality**
We mapped networks of movement across counties based on SafeGraph data (Weekly Pattern Version 2 (SafeGraph 2020a)). We used the data of visits from census block groups (CBGs) to POIs to map network movements across counties. The nodes in the network are counties; weights of edges are visits from CBGs in one county to POIs in another county. Therefore, the mapped network accounts only for the movements due to POI visits. We calculated the in-degree and out-degree centrality of each county in the mapped network and used the percentage change with respect to the first week to remove the influence of larger counties tending to have higher in-degree and out-degree centrality.

**Colocation degree centrality**
We used Facebook county-level colocation maps to calculate the colocation degree centrality of each county. Colocation maps could represent a network in which nodes are counties and edges represent the probability of population contacts between two counties. The network is undirected, and we calculated weighted colocation degree centrality as the feature reflecting mobility across counties. Also, we calculated the percentage change of the colocation degree centrality with respect to the first week.

*Features related to disease attributes*

**Reproduction number ($R_0$)**
The reproduction number ($R_0$) is an attribute of infectious diseases which estimates the number of secondary cases infected by the first case (Dietz 1993). We calculated the reproduction number of COVID-19 according to Equation 5 based on a simple epidemic transmission model (Fan et al. 2020; Ramchandani et al. 2020). The model assumes that one case would infect $R_0$ cases after a time interval $\tau$. Then $i(0)$ infected cases at the time step 0 will lead to $i(t) = i(0)R^{t/\tau}$ number of infected cases at time step t.

$$R_0 = e^{K\tau} \tag{5}$$

where $K = (\ln i(t) - \ln i(0))/t$, and we used $i = 5.1$ days for COVID-19 (Zhang et al. 2020). Furthermore, we used the percentage changes of $R_0$ with respect to the first week as the feature inputs in the model.

*Features related to social network structure*

**Social connectedness index (SCI)**
We used the social connectedness index to account for social network structures affecting epidemic transmissions (Kuchler et al. 2020). The county-level social connectedness index provided by Facebook and SCI for two counties is calculated according to Equation 6 (Facebook 2020).

$$SCI_{i,j} = \frac{FB\_Connections_{i,j}}{FB\_Users_i \times FB\_Users_j} \tag{6}$$

We can find from Equation 6 that SCI of counties $i$ and $j$ is determined based on the number of Facebook connections (i.e., friends in Facebook) between two counties divided by the number of Facebook users in two counties. SCI, therefore, reflects the strength of social connection between two counties.

We mapped a fully connected network based on the SCI. The nodes in the network are counties; edge weights are SCIs between counties. Then we calculated weighted degree centrality of each county and the feature inputted in the model.

## 2. Model related supplemental information

We fitted five random forest classifier models each week. The dependent variable for the five models were the five classifications of the weekly new confirmed CPP (cases per 100,000 population) of each county. For each week, we put counties with zero CPP in classification 0. Then, according Equation 1, the remaining counties were evenly divided into four classifications (i.e., classifications 1 through 4), with each classification having one of total four quantiles of CPP. Figure S1 illustrates the histograms of five classifications for five random-forest models in each week.

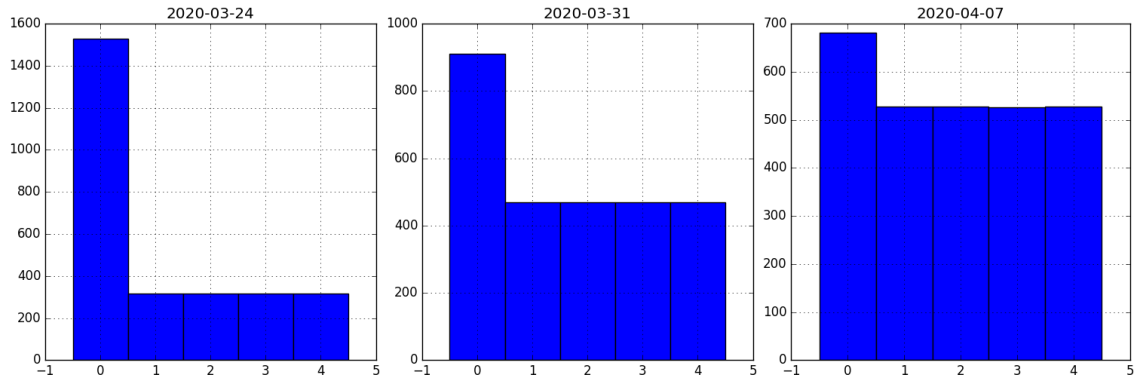

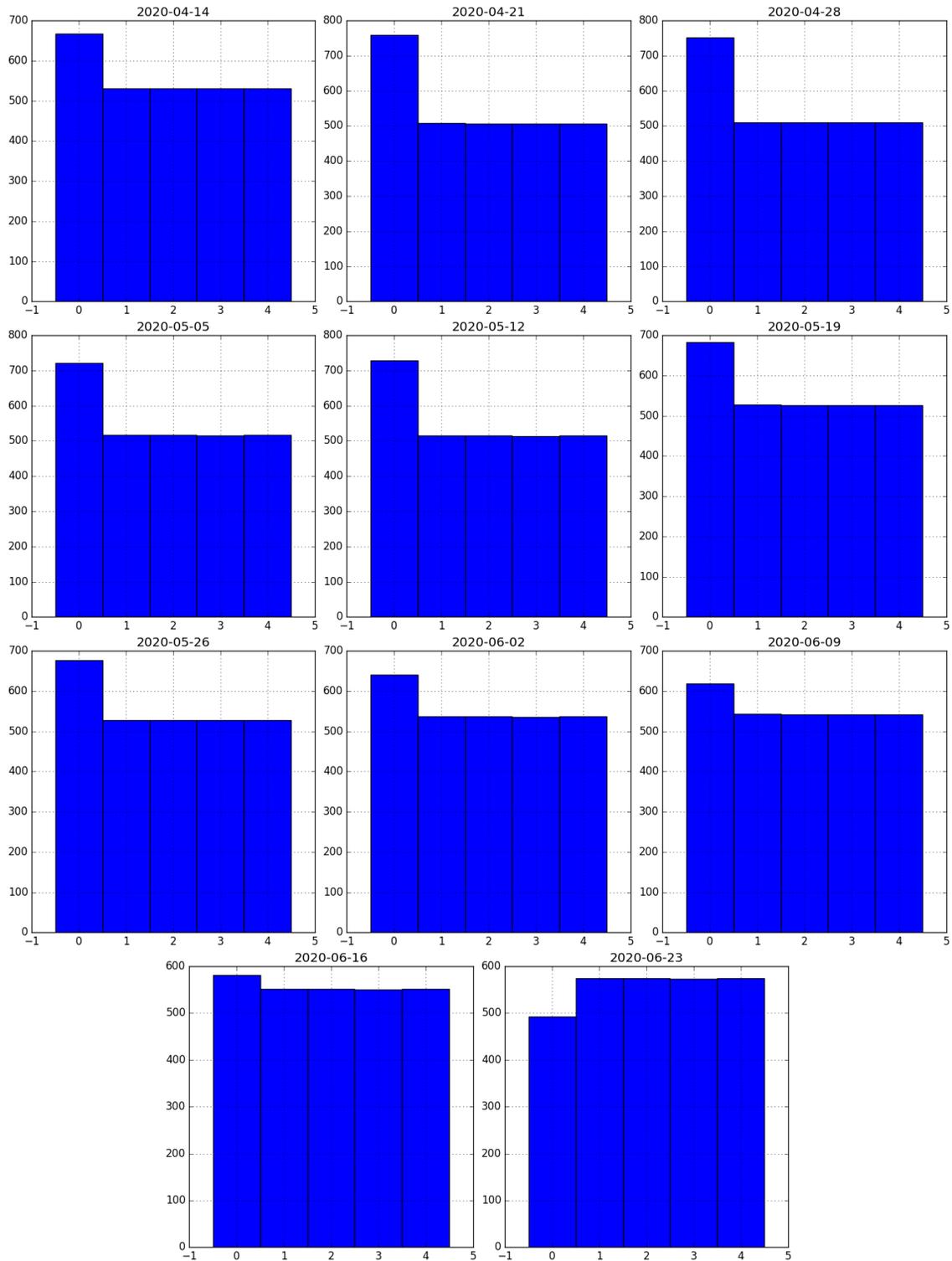
**Figure S1.** Histogram of five classifications for each week

Figures S2 illustrates the test accuracy of each model for each week. Note that random forests use the training set to determine feature importance based on reduction of Gini importance. This means that Gini importance-based features showed the extent to which the feature could well split the training set. The accuracy of the training set reached 0.99 for each model. We used ten cross-validations; the importance of each feature is the mean of ten rounds. The output accuracy is also the mean of ten rounds.

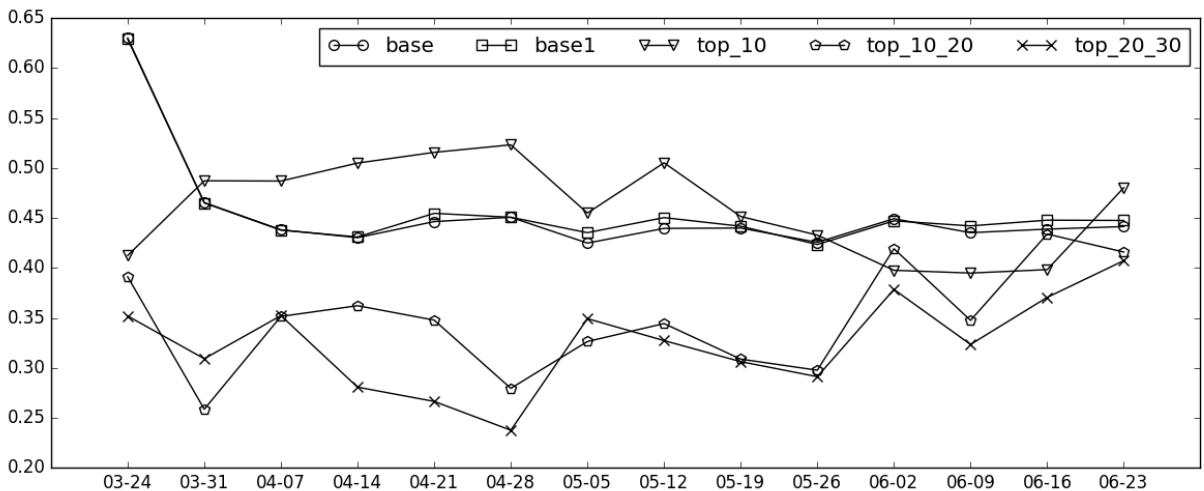

**Figure S2.** Test accuracy of each model, base: model does not include population density, base1: model including population density

**REFERENCES**
Ahmadi, M., Sharifi, A., Dorosti, S., Jafarzadeh Ghoushchi, S., and Ghanbari, N. (2020). "Investigation of effective climatology parameters on COVID-19 outbreak in Iran." *Science of the Total Environment*, 729.
Centers for Disease Control and Prevention. (2020). "County Level Social Vulnerability Index 2018." *Centers for Disease Control and Prevention*, <https://data.cdc.gov/Health-Statistics/Social-Vulnerability-Index-2018-United-States-coun/48va-t53r>.
Cuebiq. (2020). "Cuebiq's COVID-19 Mobility Insights." *Cuebiq*, <https://help.cuebiq.com/hc/en-us/articles/360041285051-Cuebiq-s-COVID-19-Mobility-Insights#h_4e44ff71-27e9-4b83-977e-d18911b21817>.
Dietz, K. (1993). "The estimation of the basic reproduction number for infectious diseases." *Statistical Methods in Medical Research*, 2(1), 23–41.
Facebook. (2020). "Facebook Social Connectedness Index." *Facebook*, <https://dataforgood.fb.com/docs/social-connectedness-index-methodology/>.
Fan, C., Lee, S., Yang, Y., Oztekin, B., Li, Q., and Mostafavi, A. (2020). "Effects of Population Co-location Reduction on Cross-county Transmission Risk of COVID-19 in the United States." *arXiv preprint arXiv:2006.01054*.
Kuchler, T., Russel, D., and Stroebel, J. (2020). "The geographic spread of COVID-19 correlates with structure of social networks as measured by Facebook." *arXiv e-prints: 2004.03055*, National Bureau of Economic Research.
Louail, T., Lenormand, M., Picornell, M., Cantú, O. G., Herranz, R., Frias-Martinez, E., Ramasco, J. J., and Barthelemy, M. (2015). "Uncovering the spatial structure of mobility networks." *Nature Communications*, 6.
Nepomuceno, M. R., Acosta, E., Alburez-Gutierrez, D., Aburtod, J. M., Aburtod, J. M., Gagnon,

A., Gagnon, A., and Turra, C. M. (2020). "Besides population age structure, health and other demographic factors can contribute to understanding the COVID-19 burden." *Proceedings of the National Academy of Sciences of the United States of America*.

Ramchandani, A., Fan, C., and Mostafavi, A. (2020). "DeepCOVIDNet: An Interpretable Deep Learning Model for Predictive Surveillance of COVID-19 Using Heterogeneous Features and Their Interactions." *arXiv preprint arXiv:2008.00115*.

Rocklöv, J., and Sjödin, H. (2020). "High population densities catalyse the spread of COVID-19." *Journal of travel medicine*.

SafeGraph. (2020a). "Weekly Pattern Version 2." *SafeGraph*, <https://docs.safegraph.com/docs/weekly-patterns>.

SafeGraph. (2020b). "Social Distancing Metrics." *SafeGraph*, <https://docs.safegraph.com/docs/social-distancing-metrics>.

Sarmadi, M., Marufi, N., and Kazemi Moghaddam, V. (2020). "Association of COVID-19 global distribution and environmental and demographic factors: An updated three-month study." *Environmental Research*, 188.

Surgo Foundation. (2020). "The COVID-19 Community Vulnerability Index." *Surgo Foundation*, <https://precisionforcovid.org/ccvi>.

U.S. Department of Commerce. (2020). "County Level GDP." *U.S. Department of Commerce*, <https://apps.bea.gov/regional/downloadzip.cfm>.

Wright, A. L., Sonin, K., Driscoll, J., and Wilson, J. (2020). "Poverty and Economic Dislocation Reduce Compliance with COVID-19 Shelter-in-Place Protocols." *SSRN Electronic Journal*.

Zhang, J., Litvinova, M., Liang, Y., Wang, Y., Wang, W., Zhao, S., Wu, Q., Merler, S., Viboud, C., Vespignani, A., Ajelli, M., and Yu, H. (2020). "Changes in contact patterns shape the dynamics of the COVID-19 outbreak in China." *Science*, 368(6498), 1481–1486.